\documentclass[11pt,a4paper]{article}
\usepackage[utf8]{inputenc}
\usepackage[T1]{fontenc}
\usepackage[english]{babel}
\usepackage{amsmath}
\usepackage{amsfonts}
\usepackage{amssymb}
\usepackage{graphicx}
\usepackage[textsize=tiny,colorinlistoftodos]{todonotes}
\usepackage{hyperref}
\usepackage{enumitem}
\usepackage{authblk}
\usepackage{lscape}
\usepackage{xcolor}

\usepackage[backend=biber]{biblatex}
\usepackage{filecontents}

\usepackage{csquotes}
\usepackage{caption}
\usepackage{subcaption}
\usepackage{rotating}
\usepackage{comment}
\usepackage{longtable}  
\usepackage{pdflscape}

\usepackage{lineno}
\usepackage[a4paper,left=2cm,right=2cm,top=2.5cm,bottom=2.5cm]{geometry}
\usepackage{setspace}
\raggedright
\title{Data needs for integrated economic-epidemiological models of {\color{black}pandemic mitigation} policies}

\author[1,*]{David J Haw}
\affil[1]{MRC Centre for Global Infectious Disease Analysis \& WHO Collaborating Centre for Infectious Disease Modelling, Abdul Latif Jameel Institute for Disease and Emergency Analytics, Imperial College London, United Kingdom}
\author[1]{Christian Morgenstern}
\author[2, 1]{Giovanni Forchini}
\author[1]{Rob Johnson}
\affil[2]{USBE, Ume\aa{ }   Universitet \\ SE-901 87     Ume\aa{ }  Sweden.} 
\author[1]{Patrick Doohan}
\author[4, 5]{Peter C Smith}
\affil[5]{Centre for Health Economics, University of York, United Kingdom.}
\author[1]{Katharina D Hauck}
\affil[*]{Correspondence: d.haw@imperial.ac.uk}

\begin{document}
\maketitle

\begin{abstract}
	The COVID-19 pandemic \textcolor{black}{and the mitigation policies implemented in response to it have} resulted in economic losses worldwide. Attempts to understand the relationship between economics and epidemiology has lead to a new generation of integrated mathematical models. The data needs for these models transcend those of the individual fields, especially where human interaction patterns are closely linked with economic activity. In this article, we reflect upon modelling efforts to date, discussing the data needs that they have identified, both for understanding the consequences of the pandemic and policy responses to it through analysis of historic data and for the further development of this new and exciting interdisciplinary field. 
\end{abstract}

\section*{Introduction}

Economic disruption during the COVID-19 pandemic has led to a wealth of mathematical models that integrate the fields of economics and epidemiology, {\color{black}e.g. \cite{Haw2022, Eichenbaum2020, Acemoglu2020, Cakmakli2020, Pichler2020}. A common theme across such studies is the association of an economic cost to the efforts made in mitigating the COVID-19 pandemic.} {\color{black} Models analyse the impact of policy interventions and/or behavioural effects on the dynamics of infection transmission and the economy of a particular location.}

\begin{figure}[h]
\begin{center}
\begin{tikzpicture}[scale=1,transform shape,thick,
main node/.style={rectangle, draw=black, very thick, fill=blue!20, text width=7em,align=center, rounded corners, minimum height=2em},
second node/.style={rectangle, draw=black, thick, fill=blue!20 ,align=center, minimum height=2em},
third node/.style={rectangle, draw=black, thick, fill=blue!20, text width=6em,align=center, minimum height=2em}]
                \draw[teal,thick,fill=teal!10] (-2.5,7) rectangle (10.5,5);
                \draw[yellow,thick,fill=yellow!10] (.3,.5) rectangle (7.7,-3.5);
                \draw[red,thick,fill=red!10] (-2.5,4) rectangle (10.5,1);
                \node[text=black,font=\small\sffamily] (1) at (-1.1,6.7) {OBSERVATIONS};
                \node[text=black,font=\small\sffamily] (1) at (1.5,-3.2) {EPI MODEL};
                \node[text=black,font=\small\sffamily] (1) at (9.3,1.3) {ECON MODEL};
                
				\node[second node,fill=white,minimum width=25pt] (1) at (1.5,-.5) {$H(\tau-1)$};
				\node[second node,fill=white,minimum width=25pt] (2) at (4,-.5) {$H(\tau)$};
				\node[second node,fill=white,minimum width=25pt] (3) at (6.5,-.5) {$H(\tau+1)$};
				\node[second node,fill=white,minimum width=25pt] (11) at (4,-2) {$\lambda(t)$};
				\node[second node,fill=white,minimum width=25pt] (4) at (8.5,3) {Economy $(\tau)$};
				\node[second node,fill=white,minimum width=25pt] (5) at (4,2.2) {Behaviour $(\tau)$};
				\node[second node,fill=white,minimum width=25pt] (6) at (-.5,3) {Policy $(\tau)$};
				\node[third node,fill=white,minimum width=25pt] (7) at (-1,6) {Stringency};
				\node[third node,fill=white,minimum width=25pt] (8) at (2,6) {Mobility};
				\node[third node,fill=white,minimum width=25pt] (9) at (6,6) {Transactions};
				\node[third node,fill=white,minimum width=25pt] (10) at (9,6) {GVA $(\tau)$};
				
				\path[every node/.style={font=\sffamily\small}]
				(1) edge [line width=1pt,->,ultra thick,dashed] node[auto] {} (2)
				(2) edge [line width=1pt,->,ultra thick,dashed] node[auto] {} (3)
				(6) edge [line width=1pt,->,ultra thick] node[auto] {} (4)
				(6) edge [line width=1pt,->,ultra thick] node[auto] {} (5)
				(6) edge [line width=1pt,->,line width=3pt,bend right=80] node[below left] {} (11)
				(5) edge [line width=1pt,->,ultra thick] node[auto] {} (4)
				(5) edge [line width=1pt,->,line width=3pt,bend left=40] node[above right] {} (11)
				(2) edge [line width=1pt,->,line width=3pt] node[auto] {} (5)
				(1) edge [line width=1pt,->,line width=3pt] node[auto] {} (6)
				(6) edge [line width=1pt,->,ultra thick,dashed] node[auto] {} (7)
				(5) edge [line width=1pt,->,ultra thick,dashed] node[auto] {} (8)
				(5) edge [line width=1pt,->,ultra thick,dashed] node[auto] {} (9)
				(4) edge [line width=1pt,->,ultra thick,dashed] node[below right] {(delay)} (10)
				(11) edge [line width=1pt,->,ultra thick] node[auto] {} (3);
\end{tikzpicture}
\end{center}
\caption{Schematic of integrated epi/econ models: epidemiological variables are grouped at the bottom (yellow), latent economic variables displayed in the centre of the figure (red), and current economic data at the top (green). For simplicity we exclude core epidemiological data (e.g. hospitalisations and deaths) from this diagram. The letter $\tau$ indexes time intervals, typically of one month in duration. GVA in period $\tau$ is reported in period $\tau+1$. In the epidemiological model, $H(\tau)$ denotes total hospital occupancy in period $\tau$, and $\lambda(t)$ the (continuously varying) Force Of Infection (FOI). Solid arrows denote dependencies between variables. Bold arrows show the causal relationships between economic and epidemiological phenomena. Dashed arrows show the manifestation of latent economic variables in current data sources, and indicate the progression of time periods in the epidemiological part of the model.}
\label{fig:flowchart}
\end{figure}
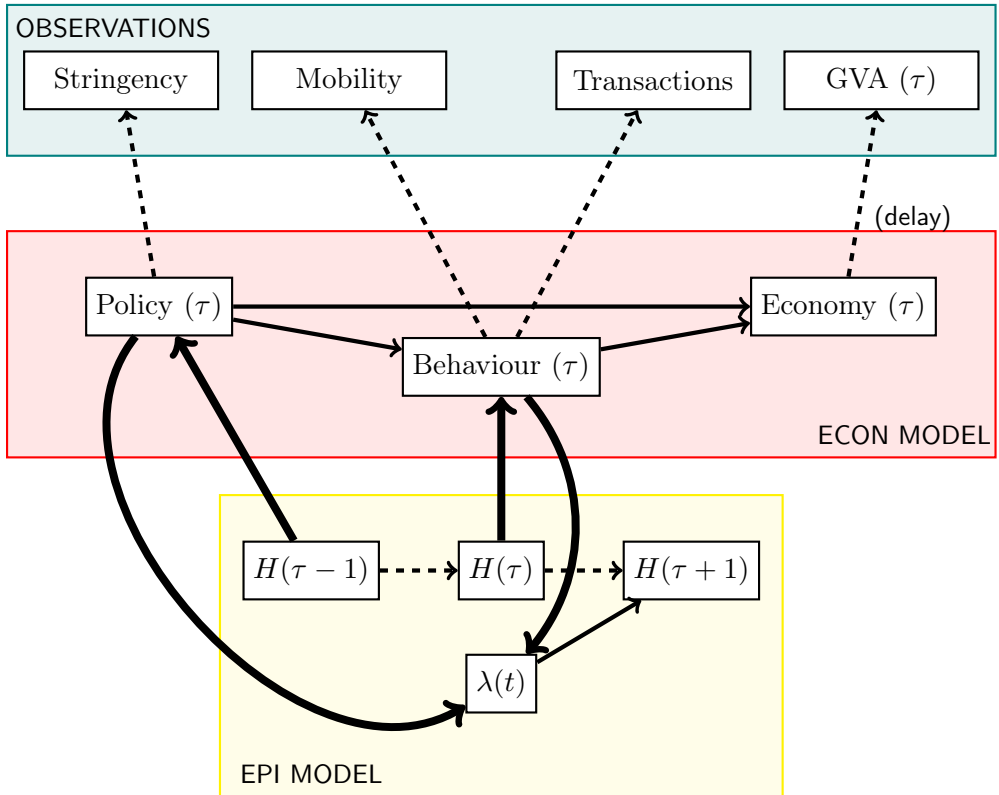

In this article, we discuss the data needs of integrated models, focusing on studies that model the impact of lockdowns and/or individuals' behaviour: mandated closures of non-essential business activity that impact on both disease transmission and economic activity or output. In these models, force-of-infection is usually affected by economic closures, i.e. mandated closures that reduce transmission and also manifest in economic variables. {\color{black}The mechanistic relationship between economic closures and transmission hazard yields mathematical models that project the impact of an economic decision on the spread of a disease, or vice versa. Figure~\ref{fig:flowchart} illustrates the causal relationship between integrated model variables and their manifestation in current data sources reported longitudinally throughout the COVID-19 pandemic. The latent variables "policy" and "behaviour" denote any degree of mandated and voluntary mitigation respectively. We define an integrated  model as one in which at least one of the bold arrows is accounted for, i.e. an explicit causal relationship between an economic variable and an epidemiological one. The exact manifestation of this relationship is dependent on the question being asked, though questions regarding the interplay between disease dynamics and economic activity are increasingly prevalent. Table~\ref{table:currentData} contains a breakdown of the data sources referenced in the figure, together with other existing sources discussed in this manuscript. With many new, large data sources emerging in recent times, both public and private, any attempt to classify data for integrated models would be quickly outdated. To this end, we use the term "data needs" to identify data gaps that go beyond data that do not exist, to encompass inconsistency of data across countries or institutions, imprecision of survey questions, and public in-accessibility of data.}


Motivated almost entirely by the COVID-19 pandemic, reflection on data sources since December 2019 allows us to identify data needs for modelling the macro-economic impact of a pandemic on the economy, and the micro-economic impact on the behaviour and decisions of individual consumers and businesses. Our discussion assumes a readership familiar with the basic underlying models. Many of our observations are rooted in the integrated model DAEDALUS \cite{Haw2022}, though broader literature searches of relevant models \cite{brodeur2021} have shown these observations to be somewhat universal. {\color{black}Many of the existing data sets identified are unique to the United Kingdom, though our focus is on the nature of the variables and the scope of the study, rather than the population from which such data was obtained.}

{\color{black}The sections in this article highlight data needs when modelling the relationship between economic activity and the transmission of an infectious disease in a population. We begin by breaking down the interface between economics and epidemiology, identifying the data heterogeneities required for alignment of models. 
We then provide a brief introduction to economic models for a readership of epidemiologists, before discussing a crucial metric of economic output, namely Gross Value Added (GVA), and the way it shapes economic modelling. One particularly difficult heterogeneity to fully encapsulate is that of geographical space, which we discuss in detail. We then explain the relationship between economic activity and physical contact out of the workplace, and the broader issue of changes in individual behaviour that give rise to changes in consumer demand. Our discussion of models and data sources will become increasingly granular, aiming to illustrate the varying levels of model complexity that researchers may wish to consider. As always, complexity is dependent on the question we are asking, but may also be limited by the data sources that are available to inform a model. A broad research question to keep in mind throughout our discussion is the following: "how can we characterise an optimal mitigation strategy that accounts for economic impact, and how could we find such a strategy in a future outbreak?" We conclude with a table of notable known data sources that illustrate the data needs discussed, briefly summarising their contents and limitations.}

\section*{The interface of economics and epidemiology}

Economics meets epidemiology where people meet people. Much of the heterogeneity in epidemiological models is born of the differences in contact rates (people per day) with respect to demographic phenomena, most notably age \cite{Mossong2008, Prem2017} and space \cite{Haw2019}. Such heterogeneities are key in replicating observed data in mathematical models \cite{Gog2014}. The relationship between economics and epidemiology that we focus on can be summarised as follows: partial economic closures reduce the active workforce interacting with each other and with customers, causing a reduction in contacts between people, reducing transmission risk. This reduced workforce also yields a reduction in production of goods and services. Likewise, the incentive to reduce contact with others causes a reduction in demand as fewer people choose to spend money out of the home, and labour force is reduced due to direct impact of infection in workers. Crucially, changes in contact patterns occur not just between workers of the same sector, but also between workers and customers (of other sectors or in the community), or between members of the community (patrons of bars/restaurants etc.). Sectors of particular note are: education (in which pupils/students are the customer), hospitality/entertainment, and travel/hotels \cite{Haw2022}. The heterogeneity necessary for a dynamic macroeconomic integrated model is therefore a mapping between economic activity and human contact. This mapping can work in either direction: mandated closures yield reductions in contact, epidemiological outcomes trigger mandated closures, or perceived threat due to an outbreak yields changes in behaviour and therefore in economic activity (c.f. the interactions between policy, behaviour, FOI $\lambda(t)$ and hospital occupancy, as an indicator of threat, in figure~\ref{fig:flowchart}). As we see a wealth of data emerge including simple metrics for behavioural change (mobility \cite{GoogleMobility}, mask wearing \cite{Riley2020} etc.), we see a clear modelling need whereby human behaviour is an explicit component of Force-Of-Infection. 

Estimates of sector-stratified contact rates in the literature are rare \cite{Beraud2015, Danon2013}, with little sector disaggregation and sparse responses in many sectors. These studies have, however, been used as the basis of many economically stratified models in the wake of COVID-19 \cite{Janiak2021, Hill2021}. In order to study the relationship between economic variables (e.g. GVA, discussed below) and contact rates, we require disaggregate data for both phenomena simultaneously, and at different points during the recent pandemic, i.e. under different regulations and different patterns of incidence/prevalence, {\color{black}and from a part of the world relevant to the country being modelled.} 

{\color{black}Intrinsic to sector- or workplace-stratified contact data is the proportion of workers working remotely, and corresponding changes in productivity. We must also differentiate between capacity for remote work should this be mandated, and baseline remote working rates, for which pre-COVID data will be long outdated. The latter represents a new workplace configuration that serves as the initial condition for a future outbreak, and may still not have been established in much of the world. Remote working data is therefore an ongoing need when modelling human workplace contact.}

{\color{black}Human contact is a result of physical presence in the workplace, whereas productivity stems from active workforce. A key data need here is changes in number of workers and proportion of time spent working remotely. The dominant data need identified here is a volume of contact surveys throughout the recent pandemic. A wealth of survey data identifies only some formal sectors, such as healthcare, education and hospitality, with a focus more on the nature of the work than its interaction with rest of the economy \cite{Riley2020}. Such data in fact incentivises a modelling need, i.e. disaggregation by work type. }

{\color{black}Assortative mixing by age is crucial to encapsulate the dynamics of airborne pathogen transmission \cite{RohaniBook}. In modelling terms, when adding further heterogeneities such as economic sector, the populations of these additional compartments may be non-uniform in age. There is therefore the potential for a change in sector-specific infection rates. For example, if customer-facing roles in hospitality are dominated by younger workers who are also densely connected outside of the workplace, then we will observe a more rapid initial spread of a pathogen within this sector. In order to observe such effects in a model, we require the age-stratification of our populations in all other compartments.}

{\color{black}Evidence from the COVID-19 pandemic has shown heterogeneity in infection risk and severity with respect to socio-economic status \cite{Lo2021, Mena2021}. Integrated modelling is a key tool in identifying how this relates to the workplace setting. In order to fully characterise the impact of stratified economic closures, we must explore the differential impact on different socio-economic groups. This may manifest in terms of workforce composition, household size, social contact patterns, ability to work remotely, ability to afford childcare, and access to medical care. There are two clearly identifiable missing links in current data: a mapping from labour force in formal economic sectors to information that is crucial for epidemiological models, most notably infection risk and contact patterns by occupation, and information on age, socio-economic status and employment type with respect to customer exchange. Reported contact patterns with respect to such disaggregation throughout different stages of economic closure would be gold standard.}

In the education sector, it is the pupils and students that are the consumers, and it is between these consumers that much of disease transmission occurs when the sector is active. For example, in the UK, efforts in online teaching and bubbling of classes or year groups have reduced such contacts and have been somewhat dynamic in response to outbreaks within a given institution. {\color{black}Furthermore, school closures create a demand in childcare that may impact the workforce in other sectors. Estimated loss in labour force and GDP due to school closures has been inferred from a combination of school closure data and household survey data \cite{Raitzer2020}, though sources used predate the COVID-19 pandemic and so estimates do not account for patterns that arise due to additional closures. Refined estimates are therefore reliant on longitudinal variants of these data sets throughout the pandemic.}

The calibration of integrated models to a disease outbreak requires epidemiological data - typically hospitalisations or deaths, owing to the more consistent reporting over time. Such consistency is a broader epidemiological problem and therefore is left to other articles in this issue. Moreover, though the disaggregation of the population into economic sector is an important heterogeneity for integrated epi-econ models, {\color{black}this heterogeneity is not necessarily directly manifest in epidemiological data resulting from changes in contact patterns: in the case of COVID-19,} hospitalisations are observed in the older and more vulnerable population, for which there is little motivation for economic stratification. 

{\color{black}We identify the following data needs for the description of heterogeneity in the interface of economics and epidemiology:}

{\color{black}
\begin{itemize}
    \item Proportion of employees and proportion of time spent working remotely, alongside any measured changes in productivity (though the latter may be inferred from sectoral outputs)
    \item Longitudinal contact survey data regarding number and nature of contacts, stratified by employment status, economic sector and type of workplace, age and socio-economic status
    \item Childcare protocols during periods of school closure (household survey level; school openings for children of essential workers; socio-economic status of schools)
    \item Daily reported numbers of school attendance and absence by reason for absence (available for the UK \cite{ONSSIS})
    \item Consistent longitudinal reporting of an epidemiological variables such as hospitalisations
\end{itemize}
}

\section*{Economic Models}

Models of the economy can be broadly classified into two domains: macro-economic models (describing the economy as a whole, or highly aggregated by economic sector - akin to compartmental models in epidemiology) and micro-economic models (describing individual economic agents such as consumers, workers or firms - akin to agent-based models (ABMs) or individual contact network models in epidemiology). Unlike the study of communicable disease, where SIR- or SIS-like dynamics characterised by force-of-infection (FOI) form the basis of all models, there is little consensus \textcolor{black}{on the essential elements of an epi-econ model. Most studies focus on a single, closed economy, though some model transmission and/or economic exchange between heterogeneous regions of a country, or a country and its main trading partners. \cite{deb2021},\cite{verschuur2021}.}

Macro-economic models focus on aggregate economic outcomes, such as the magnitude and nature of perturbations to the economy \textcolor{black}{e.g. \cite{Acemoglu2020}}. These macro-economic models of the economy include input-output models (IO models), general equilibrium models (GEMs), growth models and aggregate demand – aggregate supply models (AD-AS models). Data needs for such models are one-time surveys or time series of macro-economic outcomes: gross domestic product (GDP), gross value added (GVA) by economic sector, labour force, labour productivity, capital, consumption, savings, investments, and input-output tables (IO tables, which describe supply chains and the interdependencies between economic sectors). Though there is a clear mapping between model variables and data variables in macroeconomics, data sources such as IO tables are typically available only retrospectively, and with substantial temporal aggregation (periods of one year). 

Micro-economic models study the behaviour of economic agents and how they change as a result of exogenous shocks e.g. \cite{Eichenbaum2020}. Such models attempt to model the factors underlying changes in goods and services produced, or the changes in the supply of labour or model of working (office working vs. remote working), changes in consumption, or changes in productivity. Micro studies may allow for heterogeneity across agents, but more commonly are formulated as Representative Agent Models (RAMs). They assume that there is a limited number of classes of agents that all behave identically (in the same class). This typically involves the modelling of preferences to describe the typical behaviour of an individual in each class, assuming homogeneity in behaviour within the class. 

{\color{black}Some integrated models incorporate an endogenous behaviour change of individuals via a utility function that usually incorporates perceived risk or cost of infection, aside from other benefits and costs, which impacts consumption and/or labour supply \cite{Eichenbaum2020}. This association is informed by individuals’ preferences and internal trade-offs between the benefits of consumption and working, and the costs of infection. Integrated epi-econ models with endogenous behaviour often incorporate risk of infection via a simple SIR-like outbreak model. Such models, however, remain theoretical due to the abstract nature of individual preferences. Data inputs include consumption and hours worked \cite{Eichenbaum2020}, though no consensus on the arguments and the functional form of utility functions is apparent in the literature. Indicators for such behavioural variables have been quantified on an international scale \cite{ImperialYouGov}, though data needs for the calibration of utility functions may also depend on the choice of function in a given model.}

\section*{GVA and epi-econ modelling}

When modelling a heterogeneous economy, sector-stratified production data is advantageous over aggregate measures such as national GDP. GVA by economic sector is readily available for many countries and represents the difference between output and intermediate inputs (goods and services purchased from other sectors) such that total GVA across all sectors yields GDP. A GVA of £1bn could therefore correspond to an output of £1.5bn with intermediate inputs of £500m, or an output of £2bn with intermediate inputs of £1bn.

Economic closures affect both supply (workers are sent home, with an associated drop in productivity and economic output), and demand (there is less opportunity for consumption). This affects particularly services, or goods where market exchange cannot be online. Moreover, with knowledge of an impending lockdown, companies in a specific sector may reduce output expecting a reduced demand for consumption in order to avoid over-production. It is also feasible that the remote production of goods and services became more efficient as the pandemic progressed. For example, the initial lockdown in the UK (March 2020) showed notably lower GVA across many sectors compared to the November 2020 lockdown, despite similar mandated closures. Changes in investment are often considered negligible in early months of a pandemic, but they become increasingly important when studying long-term projections after a dramatic perturbation to the economy. 

Current reporting of GVA by economic sector presents two notable immediate limitations: a delay in reporting means that the current state of the economy often cannot be used for calibration or projections, and the temporal resolution of such data is at best aggregated by month, showing only coarse dynamic trends on the timescale of a pandemic. Moreover, thresholds of time intervals align to calendar months and so do not typically map to changes in mandated economic closures. The monthly Office for National Statistics (ONS) surveys of economic activity in the UK on which GVA is based \cite{ONSGVA} are regularly incomplete upon first publication and are usually updated as more data become available. This results in the need to re-calibrate models to the entire pandemic on a regular basis in order to accurately account for economic variables. {\color{black}The nature of GVA as an aggregate variable renders high temporal resolution problematic, though weekly or fortnightly data is not infeasible. More important is the need for nowcasts, i.e. projections that provide estimates for current economic activity, to fill the void created by reporting delay, which provide a methodological solution to this data gap. This is already available for national GDP data across the OECD countries \cite{oecdeco} and an extension to other economic time series would be very welcome.}

GVA data, like much economic data, is disaggregated by country and economic sector. The sector stratifications are, however, not uniform between countries. A baseline 117 sector configuration, when presented, is universal, though data is typically presented in aggregate form that is suited to the economy of a given country. This is problematic when subdivisions of the reported sectors pose notably different risk of disease transmission, such as the subdivision of the retail sector into “online” and “in-person”. {\color{black}When modelling at supra-national level, whether multiple interacting open economies or simply comparison between closed economies of different countries, a simple data need involves a standardised view of a single economy, mediated by a global organisation. Even a coarse but unified description would be of immediate appeal. A significant amount of excellent work has been already done by the OECD \cite{oecd_data}, IMF \cite{imf_data} and World Bank \cite{wb_data} in this space. A broader overview is provided by \cite{eco_databases}.}

{\color{black}Additional macroeconomic variables used in some integrated models include labour force (number of workers), labour supply (number of potential workers, including those furloughed and unemployed) and workforce productivity (ratio of GVA to labour force). Intuitively, reduced economic activity in a sector corresponds to a reduced labour force due to mandated closures or furloughs, together with increased remote working, and hence fewer contacts in the workplace.} 
It may also be the case that total labour force is reduced due to actual or anticipated reductions in demand, and an associated reduction in the demand for labour {\color{black}Also, the relationship between productivity and labour force might be nonlinear at the business level \cite{Feenstra2015, Serrasqueiro2006}, and may differ by sector. For example, a loss of labour force may be to some extent counteracted by increased productivity in remaining workers}.

The contact patterns experienced between workers is dependent on the worker population. Furthermore, the workforce requirements in a given sector in terms of skill set or training as well as overall number may change due to mandated closures. For example, the hospitality sector may stay active for home delivery from restaurants, thus requiring kitchen staff and delivery staff, but not service staff. Workforce composition at different stages of closure may be informed by labour earnings as a fraction of GVA \cite{ONSlabour}, though this data yields no explicit information regarding contact patterns. Crucial to understanding this relationship is the linkage of labour earnings with contact survey data discussed above. 

In many low- and middle-income countries (LMICs), a considerable portion of economic activity is attributed to the “informal sector” (or the “grey economy”) and the unregulated labour market, i.e. trade that is not taxed or monitored by the government, and day labourers that are effectively self-employed. As a result, such activity does not appear in official data, and is difficult to quantify empirically \cite{Octavia2020, CM2020}. Since it is difficult to subject the unregulated informal sector to mandated closure, reduction in transmission cannot be achieved via sending workers home. Rather, reductions in activity of the informal sector may result from reductions in demand, either as a result of mandated closures in formal sectors or behaviour change due perceived threat of infection \cite{Adom2020}. We also face the possibility of increased informal activity as activity in formal sectors is limited, though evidence shows this not to be the case \cite{Komin2020}. The case for financial support of the informal sector in LMICs is made clear in the literature \cite{Ezimma2020, Komin2020}, though much of the data is qualitative. A quantitative approach requires not just the collection of new data, but a methodology for identifying informal supply and demand. 

School closures have long been known as an effective infection control measure \cite{Chao2010}. Indeed, contact rates in children are often stratified with respect to school terms \cite{Prem2017}, and from a modelling perspective these contact rates can be seen to drive an epidemic trajectory \cite{Haw2021p}. However, in economic data the sector labelled “education” often aggregates output across primary, secondary and further education, spanning an age range of 4-21+, and includes both students and staff. The variability in contact patterns across this sector, and indeed compliance to NPIs such as distancing and mask wearing, does not require explanation. {\color{black}Moreover, 
the economic value of education is usually under-estimated in national accounts because monetary valuations of educational output that relies on input prices or fees paid do not reflect the true value of education. While efforts are made in some national accounts to correct for the undervaluation, the benefits of better education that manifest over the lifetime of individuals in terms of higher incomes and better educational opportunity are difficult to estimate and fully account for \cite{Johnson2022r}. Focusing on the GVA of an aggregate education sector therefore tends to underestimate the long-term cost of school closures. An adjustment to the GVA contribution from this sector is therefore crucial in quantifying the true economic impact of school closures. Measures of such components are available over the COVID-19 pandemic in the UK \cite{ONSSIS}}.

From this discussion we identify the following data needs as developments beyond current GVA reporting:

\begin{itemize}
    \item Separation of GVA data into constituent parts: output and intermediate inputs, labour earnings, labour productivity, capital investments etc.
    \item Employment data (hours worked/full-time-equivalent (FTE) labour force/furlough) aligned with GVA
    \item GVA nowcasts, with improved temporal resolution where possible
    \item Optimal disaggregation of the economy that is consistent between countries
    \item Quantification of activity in the informal sector (low- and middle-income countries)
    \item Estimates of the true economic contribution of the education sector to the long-term welfare of an economy 
\end{itemize}

\section*{Spatial structures}

Spatial disaggregation in epidemic models is often key to replicating epidemic timing \cite{Mills2014}. Moreover, localised mitigation strategies require such heterogeneity in the underlying models. The relevance of spatial heterogeneity is driven by population density and human movement patterns, which are easily quantified owing to mobility data, e.g. from mobile phone use \cite{Grantz2020}. When developing spatially disaggregate transmission models, travel data is key to accurately modelling human movement. If we wish to add a spatial component to integrated models, where transmission risk and economic activity are causally related, then we require some spatial disaggregation of our economic variables. 

{\color{black}Economic activity data is typically not geographically disaggregated within a given country, since national accounts are maintained at national level, and single production chains regularly involve multiple locations. Exceptions include countries with a federal structure such as the the USA, where much economic data is aggregate at state level, though spatial units are typically large and do not reflect the distribution and movement of the country's population. An integrated model must therefore rely on (relatively) spatially aggregate data regarding economic activity. Including spatial heterogeneity in an integrated mathematical model requires both spatial and economic heterogeneity in the force-of-infection and hence requires an economically disaggregated description of human contact for each location. The application of such a model would be a scenario in which economic closures are geographically targeted. The resulting economic disruption would, however, be dependent on the geographical distribution of different components of supply chains within a sector. The level of heterogeneity in economic data required to support such a framework means that this particular avenue of research would certainly be complicated and in many cases would be infeasible. The extent to which movement patterns correlate with GVA contributors. Mobility data over the COVID-19 pandemic may give an indication of physical presence under different configurations of economic closure, but the concept of locally evaluated GVA remains problematic at high resolution, hence so is the problem of associating costs to localised business closures.}

Another challenge in integrated spatial modelling lies in the observation that people do not always work where they live: contributing to the economy of a city and to the community of a rural village is not uncommon in many western countries. Many models aim to study the role of household/workplace/community transmission, without further stratification of the nature of the workplace \cite{Riley2006, Haw2020net}. {\color{black}Crucially such studies typically employ agent-based modelling owing to the heterogeneity that emerges, which introduces a degree of complexity and computational demand that may prove infeasible to include in immediate outbreak response. When taking a compartmental approach, it is naïve to assume uniformity of commuting behaviour across sectors, though some simple stratifications of the sector-specific workforces may suffice to encapsulate this phenomenon, for example the urban/rural distinction for home and for the workplace, alongside the aforementioned number of workers able to work remotely, and proportion of time doing so. Though it may be possible to infer commuting patterns from economic closures and travel data combined, there are additional movement patterns that correlate strongly with economic closures. Explicit commuting data would bring clarity to this research question.}

A second manifestation of space in integrated modelling focuses not on distance but on type of location in which people spend time. Google mobility data \cite{GoogleMobility} reports percentage change in visits to locations stratified as follows: “retail/recreation”, “grocery/pharmacy”, “parks”, “transit stations”, “workplace” and “residential”. Much of the interest in the relationship between economics and epidemiology lies in the categories “retail/recreation” and “workplace”, for which no further disaggregation is available. {\color{black}Furthermore, transaction data collected by banks and credit card companies often contain limited geographical information as a physical location for the transaction is broadly reported but it is attributed to the location of the headquarters of a company rather than the location of the transaction itself \cite{Barclays, yodlee_creditcard_data}.}

{\color{black}We have highlighted the difficulty of spatially disaggregate economic modelling. We propose the following possible avenues for data sourcing in response to some specific research questions:
\begin{itemize}
    \item {\color{black}Availability of satellite image data / Geolocation data to assess contact patterns and economic activity by location (e.g. retail parks) \cite{geolocation_retail}}
    \item Understanding the transmission risk associated with commuting: decomposition of workforce and workplace data by urban/rural; survey data including commute distances, times and travel means (to inform agent-based models); remote working data (as previously discussed). 
    \item Modelling to inform targeted closures of specific premises: data should focus on the nature of a location associated with economic activity, rather than any manifestation of geographical distance, and on the presence of such locations in different geographical areas. 
    \item Mitigating animal disease outbreaks: this is an obvious exception to the cases discussed above, whereby we would aim construct a geographically explicit model of the farming supply chain. 
\end{itemize}
}

\section*{The physical nature of a transaction}

Central to an integrated epi-econ model is a description of the relationship between transactions (part of economic activity) and transmission potential (epidemiology). The latter is dependent on the nature of contact between individuals: physical proximity, duration, presence of a screen/mask, physical contact, mutual touching of equipment, online payment, delivery vs. collection of goods etc. Data regarding individual transactions \cite{Barclays} alludes to this by reporting physical presence or absence of a payment card. The limitations of this variable are as follows: a distinction is made between a contactless payment using a card and contactless mobile payment, but no distinction is made between contactless mobile payment and online payment. {\color{black}Rich data sets with transaction data, data on economic linkages, supply chains and financial transaction of physical assets exist \cite{yodlee_creditcard_data} but tend to be associated with high subscription fees.}

{\color{black}Though human} mobility data and transaction data are abundant, they are reported independently of one another and hence any relationship can only be inferred. If we wish to quantify risk aversion for application to mechanistic modelling, then we require data displaying the physical nature of transactions. We acknowledge the possibility of linking movement data with transaction data, though each is highly aggregate in different dimensions. Heterogeneity in the physical nature of a transition results in heterogeneity in contact rates and risk of transmission given contact. {\color{black} Furthermore, if certain sector closures correlate more strongly with transactions involving physical presence then economic activity in such sectors may serve as a predictor for changes epidemiologically relevant contact patterns, which is the essence of integrated modelling.}

We illustrate this by contrasting the COVID-19 pandemic with a Norovirus outbreak: the former is typically transmitted by droplets and is heavily dose dependent, whereas the latter is typically transmitted via contaminated surfaces. The role of Perspex screens and contactless payment therefore differ greatly between the two cases. For reasons of data confidentiality, it is infeasible to expect individual transaction data, though we propose aggregation by physical location such that the epidemiological distinction between different transaction environments is retained.

We propose the following candidate data needs, where some historic data may already exist privately: 
\begin{itemize}
    \item total spend in the retail/hospitality/entertainment/transport sector, disaggregated by physical nature of transaction. 
    \item Proxy variables for the latter may include the following: proportion of transactions made indoors; online payment and collection/delivery; NPIs in place on site. 
    \item Crucial to understanding the physical nature of a transaction is a longitudinal component, displaying the change in physical nature of a transaction at different stages of an outbreak (ideally including the “no outbreak” scenario).    \item {\color{black}Availability of commercial data sets to academic institutions free for research purposes}
\end{itemize}

\section*{Understanding behavioural response}

Many governments have responded throughout the recent pandemic with mandated limitations of business/economic sectors considered non-essential. Such mandates impose a limit on economic activity. The behavioural response of consumers may further reduce demand for goods and services, and the anticipation of businesses the demand for labour. For example, if restaurants are subject to a mandate of table service only, there will be reduced demand for dining (upper limit on covers), with a knock-on reduction in demand for intermediate goods and services (sourcing fewer ingredients), and employment (reduced staffing). However, perceived risk of infection by consumers may result in even greater reduction in demand. The combined effect of the mandate and behavioural effects will manifest via reduced GVA for that sector. Moreover, the behavioural component may respond to broader range of phenomena than the mandate alone: the perceived threat may depend on knowledge of prevalence or hospital occupancy, rhetoric promoted in the media, or “behavioural fatigue” \cite{Harvey2020, Cowling2010b}. Such phenomena are difficult to quantify directly due to their subjective nature, but they are manifest in metrics such as mobility, reported contacts and financial transactions. If we wish to derive mathematical description of behaviour from survey data obtained throughout the course of the COVID-19 pandemic then analysis of a large volume of these less subjective data is a good place to start, and can inform more carefully articulated questions for future surveys. 

GVA and employment alone show only the overall output, but do not help to understand which factors impact on the supply and demand-side drivers that contribute to this output. We therefore propose the following: a combination of contact survey and business activity data in hospitality venues. Crucially, observations under different sets of mandated restrictions (and necessarily different concurrent levels of prevalence/hospital occupancy) would allow for the quantification of change in contact patterns throughout the pandemic. A large volume of such data would allow modellers to explore a functional relationship between state variables and behavioural response. {\color{black}Such data does exist for the UK \cite{Riley2020, ImperialYouGov}, though much retrospective analysis is required to identify the power of these surveys, and to identify specific modifications required in the light of research questions and modelling studies that have emerged since the conception of early COVID-19 questionnaires.}

\section*{How to proceed}

Much of the available economic data regarding changes in activity throughout the COVID-19 pandemic is aggregate: in space (by country), in time (discretely by month at best) and in the output of economic goods and service under consideration. As a result, often only the aggregate outcome of economic phenomena is manifest. Disaggregation of transaction, timing and output is often needed to understand the relationship between FOI, economic activity and perceived threat. Disaggregation of the education sector into primary/secondary/further, and localised data of outbreaks and closures in schools, would allow us to refine our mechanistic description of mitigation strategies with respect to this sector.   

Where large data sets do exist regarding transactions, mobility and contact surveys, there is an opportunity for inferring parameter values that describe the interface between economics and epidemiology. Moving forward, however, explicit coupling of transaction data with the physical nature of a transaction would both confirm (or deny) and further refine estimates from more indirect methods. 

Social contact surveys are commonplace in epidemiology. However, they typically do not relay direct information regarding employment sector. “Customer facing” and “healthcare/patient facing” are standard practice \cite{Riley2020}, though there remains substantial heterogeneity in the nature of a contact within these categories. 

It is important to acknowledge the difference between a descriptive study of the relationship between economics and epidemiology throughout the COVID-19 pandemic, and the development of predictive models that require assumptions describing this relationship. With better calibrated fixed parameters, our theoretical integrated econ-epi models would gain predictive power, though in the early stages of this interdisciplinary field we are far from consensus on how such models should function. It is therefore somewhat early to describe the full data needs in the health economics of infectious diseases. 

Among the data needs we have identified are many that will never exist for past outbreaks, meaning that retrospective studies will necessarily require a greater armoury of inference methods in order to extract information about the interface of economics and epidemiology. Moreover, the baseline scenario of “no pandemic” has changed as a result of the last two years, so that a more detailed description of the pre-COVID world is beyond our reach. 

We have focused on the data needs for models in which economic mandates impact the dynamics of transmission. The converse, in which economic activity is dependent on epidemiological variables, poses the additional difficulty of quantifying behavioural response. Models must also distinguish between intervention measures that reduce some economic activity (e.g. enforced business closures, limiting capacity in public spaces), and those that do not (e.g. mask wearing, Perspex screens). The latter point is relevant in broader epidemiological studies and so we do not address it directly here. 

Prior to the emergence of COVID-19, it was widely expected that the next respiratory pandemic to hit the globe would be influenza A. Proven wrong by a pathogen with double the severity, we are reminded that the nature of future outbreaks is a known unknown, and that not all disease outbreaks require the same data. The discussion of data needs for epidemic models alone is therefore ongoing, and data for integrated models will depend on the way in which any newly emerging pathogen impacts the economy. {\color{black}We acknowledge that our discussion is centered around airborne pathogen transmission, where physical proximity can drive transmission. We propose that a more general indicator of transmission potential must differentiate between different types of contact reported, e.g. handshake vs. shared enclosed space.}

It is perhaps ambitious to envisage a mathematical model that is fully dynamic in its description of both the economy and infectious disease transmission, calibrated in real time to metrics for both phenomena. But it is perhaps also unnecessary. As modellers explore the theoretical space of integrated epi-econ models, we hope to identify precise requirements of disaggregation in data that sufficiently informs our description of the interface between economics and epidemiology, whilst remaining feasible to report in the long term. 

\begin{landscape}
\centering
\begin{longtable}{l|l|l}
\caption{Current key economic data sources for integrated modelling. Data needs, as defined in our introduction, are highlighted in the right-hand column, "limitations".}\\
\label{table:currentData}
{\bf Data type/Source} & {\bf Notes} & {\bf Limitations}\\
\hline
\endfirsthead
\multicolumn{3}{l}{\textit{Continued from previous page}}\\
\hline
{\bf Data Type/Source} & {\bf Notes} & {\bf Limitations}\\
\hline
\endhead
\hline
\multicolumn{3}{r}{\textit{Continued on next page}}\\
\endfoot
\hline
\endlastfoot
\multicolumn{3}{l}{\bf GVA}\\
\hline
Country-specific delayed reporting, & Sector-specific contributions to GDP & Reported retrospectively, by monthly interval\\
e.g. \cite{ONSGVA} (UK), \cite{ABSGVA} (Australia), \cite{BPSGVA} (Indonesia) &  & Classification of businesses into sectors is \\
 &  & inconsistent over time and between countries \\
\hline
\multicolumn{3}{l}{\bf Transactions}\\
\hline
Yodlee data analytics \cite{yodlee_creditcard_data} & Clean tagged and anonymised & Available only at high monetary cost \\
 & consumer transaction data of &  \\
  & public and private merchants & \\
\hline
E.g. Barclays UK \cite{Barclays} & Weekly indices of value and volume & Not freely available; \\
 & of credit/debit card transactions & Spatially disaggregated by location of HQ; \\
 & by group of Merchant Category Codes (MCCs) & MCCs do not align with \\
 &  & sectors of GVA/IO table configurations \\
\hline
\multicolumn{3}{l}{\bf Mobility}\\
\hline
Google COVID-19 Community Mobility Reports \cite{GoogleMobility} & Changes over time in visits to: & Baseline data is Jan/Feb 2020, \\
 & retail and recreation, supermarket and pharmacy, & giving no true pre-pandemic reference \\
 & parks, public transport, workplaces, residential; & for seasonality etc.; \\
 & Available by country/region & No distinction between Workplaces \\
\hline
\multicolumn{3}{l}{\bf Stringency}\\
\hline
COVID-19 behaviour tracker & Stringency index comprised of: & Index and constituents \\
\cite{Hale2021, Stringency} & school closures, workplace closures, & highly coarse grained at national level \\
 & cancellation of public events, &  \\
 & restrictions on public gatherings, &  \\
 & closures of public transport &  \\
 & stay-at-home requirements, &  \\
  & public information campaigns, &  \\
 & restrictions on internal movements &  \\
 & international travel controls &  \\
\hline
\multicolumn{3}{l}{\bf Sector-stratified contact rates}\\
\hline
Two known sources: \cite{Beraud2015, Danon2013} & Workforce surveys reporting & Some sample sizes are small \\
 & daily contact rates & and not representative; \\
 &  & Limited disaggregation of sectors \\
\hline
\multicolumn{3}{l}{\bf Education sector breakdown}\\
\hline
ONS COVID-19 Schools Infection Survey (SIS, \cite{ONSSIS}) & COVID-19 Swab test and survey of & Not freely available \\
 & schoolchildren; & \\
 & Includes household composition &  \\
 \hline
Attendance in education \cite{GOVschools} & School and early years settings & No linkage to household structures \\
\hline
\multicolumn{3}{l}{\bf Survey data}\\
\hline
COVID-19 behaviour tracker & Global insights on people’s behaviours & Survey discontinued. \\
(Imperial College/YouGov) & in response to COVID-19 &  \\
\hline
REACT-1 \cite{Riley2020} & Longitudinal swab and survey data; & Limited identification of employment sector; \\
 & Large number of variables reported & Not freely available; \\
 & at individual level within &  \\
 & and within households &  \\
\hline
ONS COVID-19 Infection Survey & Longitudinal swab and survey data & Not freely available; \\
(SIS, \cite{ONSCIS}) & stratified by age, region and vaccine status & Full survey contents not publically available \\
 \hline
ONS Labour Force Survey & Employment circumstances & Quarterly temporal aggregation (UK) \\
(LFS, \cite{ONSLFS}) & of the UK population &  \\
 \hline
ONS Survey on Living Conditions & Information on household resources, housing, & Only two time points: \\
(SLC, \cite{ONSSLC}) & labour, education, pensions and health & May 2021 and September 2021; \\
 &  & Two binary output variables: \\
 &  & Ability to afford unexpected expense/holiday, \\
 &  & Ability to pay usual expenses \\
 \hline
\end{longtable}
\end{landscape}

\printbibliography


\end{document}